\documentstyle[times,pramana,epsf,floats]{ias}

\newcommand{\bea}{\begin{eqnarray}}
\newcommand{\eea}{\end{eqnarray}}

\begin{document}
\mark{{Constraints on sparticle spectrum in different Supersymmetry
Breaking Models}{K. Huitu, J. Laamanen and P. N. Pandita}}
\title{Constraints on Sparticle Spectrum in different Supersymmetry
Breaking Models}

\author{Katri Huitu$^{1,2}$, Jari Laamanen$^2$ and P. N. Pandita$^3$}
\address{$^1$HEP Division, Department of Physical Sciences, University of
Helsinki, and\\
$^2$Helsinki Institute of Physics,
P. O. Box 64, FIN-00014 Helsinki, Finland\\
$^3$Department of Physics, North Eastern Hill University, Shillong 793 022, India}
\abstract{
We derive sum rules for the sparticle masses in different models of 
supersymmetry breaking. This includes the gravity mediated 
models~(SUGRA models) as well as models in which supersymmetry breaking 
terms are induced by super-Weyl anomaly~(AMSB models). 
These sum rules can help
in distinguishing between these models. In particular we obtain
an upper bound on the mass of the lightest neutralino as a 
function of the gluino mass in SUGRA and AMSB models.}
\maketitle
\section{Introduction}
Since no supersymmetric partners of the Standard Model~(SM) particles 
have been seen, supersymmetry (SUSY), if it exists, must be a
broken symmetry. Mechanisms for SUSY breaking may be classified
according to the magnitude of the gravitino mass $m_{3/2}$:
$m_{3/2}\sim 1$ TeV~(gravity mediated; SUGRA),  
$m_{3/2}\gg 1$ TeV~(anomaly mediated; AMSB)
and  $m_{3/2}\ll 1$ TeV~(gauge mediated). In SUGRA models
there are operators $\sim 1/M_P$ connecting the
hidden sector to the observable sector
which communicate the SUSY breaking.
%%%%%%%%%%%%%%%%%%%%%%%%%%%%%%%%%%%%%%%%%%%%%%%%%%%%%%%%%%%%%%%%%%%%%%%
%In this class of supersymmetry breaking
%models the
%superpotential and Kahler potential have the form
%\bea
%W & = & W_{hid} + W_{obs}, \label{sugraW}\\
%K & = & K_{hid} + K_{obs}. \label{sugraK}
%\eea
%The Kahler function is
%\bea
%G&=&K/M_{P}^2+\ln |W/M_{P}^3|^2. \label{sugraG}
%\eea
%Insertng into the F-terms of the supergravity~(SUGRA) potential
%\bea
%V_F&=&M_{P}^4e^G[G^i(G^{-1})^j_iG_j-3],
%\label{sugraV}\\
%G^i&=&\frac{\delta G}{\delta \phi_i}, \, \, \, \,
%G_i=\frac{\delta G}{\delta \phi^{*i}}, \, \, \, \,
%(G^{-1})^k_iG^j_k=\delta^j_i,
%\eea
%leads to a hidden sector SUSY breaking order parameter and a gravitino mass
%($<V_F>=0$)
%\bea
%F_i&=&-M_P^2e^{G/2}(G^{-1})^j_iG_j\\
%m_{3/2}^2&=&\frac{1}{3M_P^2}<K^i_jF_iF^{*j}>,
%\\ \nonumber
%\eea
%respectively.
%Depending on the details of the Kahler potential, terms in the expansion
%of $V_F$ then lead to soft SUSY breaking masses in the observable sector.
%Common feature of these models are $m_{3/2} \sim TeV$, and
%SUSY breaking scale in the hidden sector with
%$<F_i> \sim (3\times 10^{10} GeV)^2$.
%These models are characterised by two functions: the superpotential
%and the Kahler potential. 
In minimal supergravity models
one can choose a Kahler potential such that
squarks or sleptons have universal soft masses, and universal
soft trilinear parameters of order $m_{3/2}$. One can also choose gauge
kinetic functions so that one has universal gaugino masses $M_{1/2}$
of order $m_{3/2}$ at high energies.

On the other hand, if the soft SUSY breaking terms are  determined by the
breaking of scale invariance~\cite{rs}, then they can be written
in terms of $\beta$-functions and anomalous dimensions
in the form of relations which hold at all energies.
%\bea
%M_\lambda &=& \frac{\beta_g}{g} m_{3/2},\label{gmass}\\
%m_{i}^2 &=& -\frac 14 \left( \frac{\partial \gamma_i}{\partial
%g}\beta_g + \frac{\partial \gamma_i}{\partial y}\beta_y\right)
%m_{3/2}^2,\label{smass}\\
%A_y &=& -\frac{\beta_y}{y} m_{3/2},\label{Amass}
%\eea
An immediate consequence of such models, known as anomaly mediated 
supersymmetry breaking~(AMSB) models,  is that supersymmetry
breaking terms are completely insensitive to the physics in the ultraviolet.
However, it turns out that pure scalar mass-squared anomaly contributions
for sleptons are negative~\cite{rs}. There are a number of proposals
for solving this problem of tachyonic slepton masses~\cite{hlp}. 
The simplest of these is to 
%phenomenologically attractive
%way of parametrizing the nonanomaly mediated
%contributions to the slepton masses
add a common mass parameter $m_0$ to all the squared scalar masses,
assuming that such an addition does not reintroduce the supersymmetric
flavor problem~\cite{ggw}.

In this talk we shall discuss the sparticle sum rules in gravity
mediated models and anomaly mediated models. In particular, we shall discuss
an upper limit on the mass of the  lightest neutralino as 
a function of the gluino mass in these models.
\section{Sum rules}
In the case of minimal supersymmetric standard model~(MSSM) with
gravity mediated supersymmetry breaking, there are seven physical
sfermion masses for the first two generations which can be written
in terms of four parameters~(for a given $\tan\beta=v_2/v_1, v_1$
and $v_2$ being the vacuum expectation values of the two Higgs
doublets of MSSM). This results in three sum rules for the sparticle masses
of the first two generations~\cite{mr}, which can be used to test
the various assumptions of MSSM with gravity mediated supersymmetry breaking.
These can be written as
\bea
&&M_{\tilde d_L}^2-M_{\tilde u_L}^2 = -\cos 2\beta M_W^2, \;
M_{\tilde e_L}^2-M_{\tilde \nu}^2 = -\cos 2\beta M_W^2,\label{indep1}\\
%\eea
%and
%\bea
&&2(M_{\tilde u_R}^2 - M_{\tilde d_R}^2) +(M_{\tilde d_R}^2-M_{\tilde d_L}^2) +
(M_{\tilde e_L}^2 - M_{\tilde e_R}^2)
=  \frac{10}{3} \sin ^2\theta_W
M_Z^2 \cos 2\beta. \label{indep3}
\eea
The sum rules (\ref{indep1}), which relate the masses
of squarks and sleptons living in the same $SU(2)_L$ doublet, depend only
on the $D$-term contribution to the squark and slepton masses. They
are, thus, independent of the supersymmetry breaking model and test only the
gauge structure of the effective low energy supersymmetric model.
On the other hand the sum rule (\ref{indep3}) depends on the assumption
of a universal soft breaking mass parameter $m_0$, and is, therefore, a test
of universality of the soft scalar masses in anomaly mediated
supersymmetry breaking~(AMSB) models as well.

There are four remaining relations between the masses of the first two
generations of squarks and sleptons. In anomaly mediated models, two
of these can be used to obtain expressions  for the input parameters
$m_0$ and $m_{3/2}$ in terms of squark and slepton masses.
The remaining two equations
can then be converted to two additional sum rules. These sum rules,
which are unique to the minimal AMSB models, can be written as
\bea
&&(M_{\tilde e_L}^2 -M_{\tilde e_R}^2)+\frac 34\left(3-\frac{3}{11}
\cot^4\theta_W\right)(M_{\tilde u_R}^2-M_{\tilde d_R}^2)\nonumber\\
&&= \left[-\frac 12 + \left(\frac{17}{4}-\frac{9}{44}\cot^4\theta_W\right)
\sin^2\theta_W\right] M_Z^2\cos 2\beta\label{sum4}\\
%\eea
%and
%\bea
&&(M_{\tilde e_L}^2 -M_{\tilde e_R}^2) +\left(\frac 94
-\frac{g_2^4}{2g_3^4}
\right) (M_{\tilde u_R}^2 -M_{\tilde d_R}^2)
-\frac{3}{16}\frac{g_2^4}{g_3^4} (M_{\tilde d_R}^2-M_{\tilde e_R}^2)\nonumber\\
&&=\left[ -\frac 12+\left(\frac {17}{4}-\frac 58\frac{g_2^4}{g_3^4}\right)
\sin^2\theta_W\right]
M_Z^2\cos 2\beta.\label{sum5}
\eea
In AMSB
%anomaly mediated
models, where extra contributions to the soft squared
masses can be generated in alternative ways, one can also obtain sum rules
which can be used to distinguish these models from the
minimal AMSB model as well as from the SUGRA
%gravity mediated
models~\cite{hlp}.

\begin{figure}[h]
\leavevmode
\begin{center}
\begin{tabular}{cc}
\mbox{\epsfxsize=5cm\epsfysize=5cm\epsffile{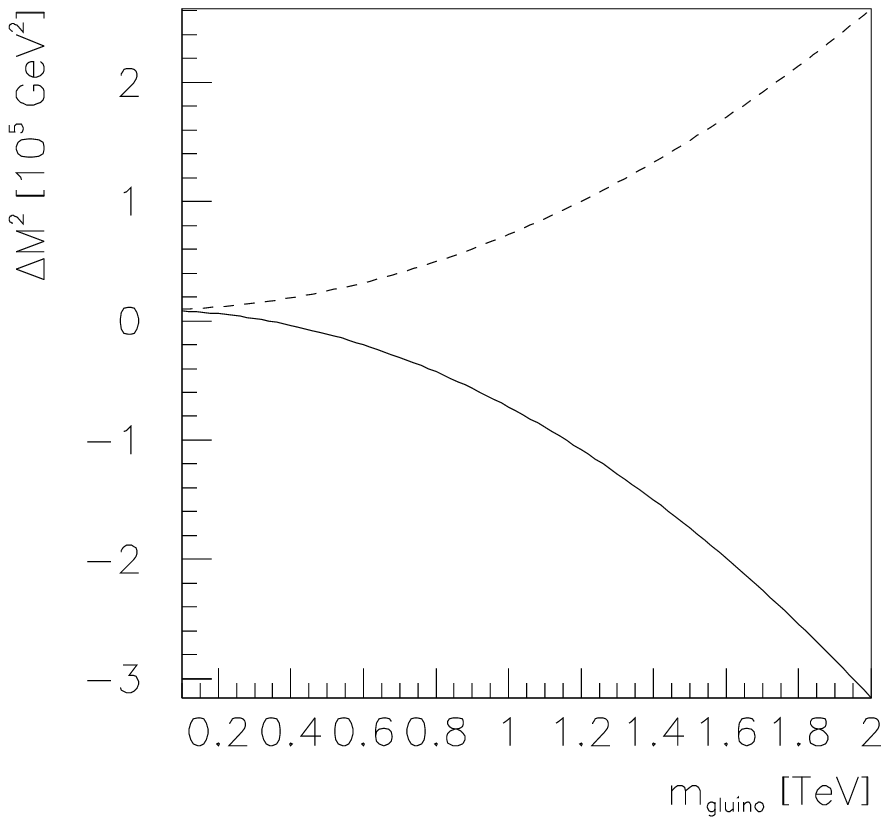}}&
%\vspace*{-1.5cm}
\mbox{%\vspace*{-1.5cm}
\epsfxsize=6.5cm\epsfysize=6.5cm\epsffile{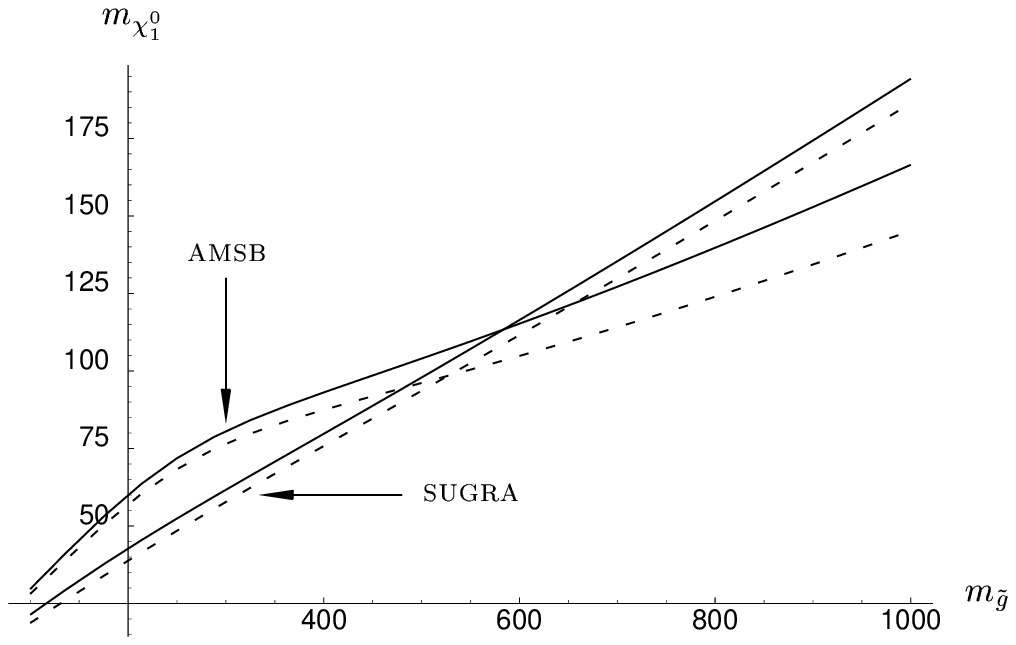}}
\end{tabular}
\end{center}
%\\mbox{
%\epsfxsize=5cm
%\epsfysize=5cm
%\leftline{\epsfbox{gaugino.eps}}
\vspace{-.25cm}
\caption{Left: The average mass difference
$\Delta m^2\equiv 2\sum M_{\chi^\pm_i}^2-\sum M_{\chi^0_i}^2$
in the AMSB models (solid line),
and in the MSSM (dashed line) as a function of the gluino mass
$m_{\tilde g}$.
Right: The upper bound on the lightest neutralino mass as a function of 
gluino mass in SUGRA
and in AMSB models. The tree level results are represented
by dashed lines and the NLO results by the solid lines.}
\label{gauginos}
\end{figure}

The  gaugino sector is same in all the models discussed in this paper.
However, in AMSB models there is a close proximity of the lightest neutralino
and chargino masses, which is a direct consequence of the soft
supersymmetry breaking gaugino mass hierarchy in these models.
Thus the winos are the lightest neutralinos and charginos, and
one would expect that the lightest chargino is only slightly heavier
than the lightest neutralino.
It is not feasible to obtain mass sum rules for the neutralinos states, since
the physical neutralino mass matrix is a
$4\times4$ matrix.
However, from the trace of neutralino and chargino mass matrices
one obtains a sum rule~\cite{hlp} which relates the average mass squared
difference:
% of the charginos and neutralinos:
\bea
%2(M_{\tilde \chi^\pm_1}^2+M_{\tilde \chi^\pm_2}^2)-(M_{\tilde \chi^0_1}^2+
%M_{\tilde \chi^0_2}^2+M_{\tilde \chi^0_3}^2+M_{\tilde
%\chi^0_4}^2)
2\sum M_{\chi^\pm_i}^2-\sum M_{\chi^0_i}^2
=\frac 19
\left[\frac{g_2^4}{g_3^4}-\left(\frac{33}{5}\right)^2
\frac{g_1^4}{g_3^4}\right]m_{\tilde g}^2 +4M_W^2-2M_Z^2,
\label{gauginosum}
\eea
We have plotted in the left panel of
Fig.~\ref{gauginos} the sum rule (\ref{gauginosum})
both in the AMSB models and the MSSM.
The average mass difference in the AMSB models is first positive,
but then quickly turns negative (solid line), while in the MSSM it is always
positive (dashed line).
Thus, this sum rule could be  one of the signatures of the AMSB type models.
\vspace*{-3cm}
%\begin{figure}[htbp]
\vspace*{-0.25cm}
Since neutralino is supposed to be the lightest supersymmetric particle
in models with $R-$parity conservation, it is of crucial importance to have a
knowledge of its mass. From the structure of the neutralino mass matrix,
we have derived an analytical upper bound on the mass of the lightest
neutralino in supersymmetric models~\cite{pnp},~\cite{hlp1}.
This is plotted in the right panel of Fig.~\ref{gauginos},
%{twobytwo},
and includes the next to leading  order radiative corrections.
We note that for most of the values of the gluino mass, the upper bound is 
different for SUGRA and AMSB models.
%\begin{figure}[htbp]
%\mbox{\epsfxsize=8.5truecm\epsfysize=8.5truecm\epsffile{sum.eps}}
%\epsfxsize=5cm
%\epsfysize=5cm
%\leftline{\epsfbox{vprans.eps}}
%\caption{ The upper bound of the lightest neutralino in SUGRA
%and in AMSB models. The tree level results are given by dashed lines and
%the NLO results by the solid lines.}
%\label{twobytwo}
%\end{figure}

%\vspace*{0.2cm}

\noindent
{\bf Acknowledgements}\\
KH and JL thank the Academy of Finland
(project number 48787) for financial support. The work of PNP is supported 
by the Council of Scientific and Industrial Research, India.


\vspace*{-0.6cm}

\begin{thebibliography}{99}
\bibitem{rs} L. Randall, R. Sundrum, Nucl. Phys. {\bf B557}, 79
(1999); G. Giudice et al., JHEP, {\bf 9812}, 027 (1998);
J.A. Bagger et al., JHEP {\bf 0004}, 009 (2000).
\bibitem{hlp} See, e.g., K. Huitu, J. Laamanen and P. N. Pandita, Phys. Rev.
{\bf D65}, 115003, (2002).
\bibitem{ggw}T. Gherghetta, G. F. Guidice, and J. D. Wells, Nucl. Phys.
{\bf B559}, 27 (1999).
\bibitem{mr} S. P. Martin and P. Ramond, Phys. Rev. {\bf D48}, 5365 (1993).
\bibitem{pnp} P. N. Pandita, Phys. Rev. {\bf D53}, 566 (1996).
\bibitem{hlp1}  K. Huitu, J. Laamanen and P. N. Pandita, (in preparation).
\end{thebibliography}
\end{document}